\documentclass[a4paper,11pt]{article}
\usepackage{pos}
\usepackage{amsmath,amscd}
\usepackage{amsfonts}
\usepackage{graphicx}
\usepackage{color}
\usepackage[normalem]{ulem}
\usepackage{hyperref}
\usepackage{enumerate}
\usepackage{mathtools}
\usepackage{slashed}
\usepackage{dsfont}

\def\tr         {{\rm  tr}}
\def\cala         {{\cal A}}

\def\calf         {{\cal F}}
\def\calg         {{\cal G}}

\def\calh         {{\cal H}}
\def\cali         {{\cal I}}

\def\call         {{\cal L}}
\def\calm         {{\cal M}}
\def\caln         {{\cal N}}

\def\calq         {{\cal Q}}

\def\cals         {{\cal S}}

\def\calz         {{\cal Z}}
\def\half{{1 \over 2}}

\def\a{\alpha}
\def\b{\beta}

\def\l{\lambda}

\def\x{\xi}

\def\o{\omega}

\def\P{\Pi}

\newcommand{\rom}[1]{\uppercase\expandafter{\romannumeral #1\relax}}

\title{Superconformal Symmetry and Index Theory}
\author[a]{Joris Raeymaekers}
\author*[a]{Canberk \c{S}anl{\i}}
\author[b,c]{Dieter Van den Bleeken\footnote[2]{Currently at Department of Meteorological and Climate Research, Royal Meteorological Institute 1180 Uccle, Belgium.}}

\affiliation[a]{CEICO, Institute of Physics of the Czech Academy of Sciences,\\  Na Slovance 2, 182 00 Prague 8, Czech Republic.}

\affiliation[b]{Physics Department, Boğaziçi University\\
	34342 Bebek / Istanbul, Turkey}
\affiliation[c]{Secondary address:\\
	Institute for Theoretical Physics, KU Leuven\\
	3001 Leuven, Belgium}

\emailAdd{joris@fzu.cz}
\emailAdd{sanli@fzu.cz}
\emailAdd{dieter.vdbl@gmail.com}

\abstract{
Formulation and supersymmetry localization of superconformal indices for $\caln=2B$ superconformal quantum mechanics are reviewed by providing a generalization to fixed point submanifolds of resolved target space geometries, and future applications to gauged scaling quivers are discussed.}
		
\FullConference{Corfu Summer Institute 2023 "School and Workshops on Elementary Particle Physics and Gravity" (CORFU2023)\\
	23 April - 6 May , and 27 August - 1 October, 2023\\
	Corfu, Greece\\}

\begin{document}
	\maketitle
	
	\section{Introduction}

Strominger and Vafa \cite{Strominger:1996sh} initiated a black hole microstate accounting programme calculating black hole entropy%\footnote{The subscript BH is a shorthand for `Bekenstein-Hawking' \cite{}.}
\begin{equation}
		S_{BH} = \frac{Ac^3}{4 G\hbar}, \label{bhent}
\end{equation}
by a count of D-brane microstates in a regime where string coupling constant is very
small. However, an explicit understanding of $AdS_2/CFT_1$ duality via \cite{Sen:2008yk}
\begin{equation}
	e^{S_{BH}(\vec{q})} = \Omega(\vec{q}),	
\end{equation} where $\Omega(\vec{q})$ denotes the degeneracy of ground states carrying charge $\vec{q}$ in the dual $CFT_1$, continues to remain as the least understood (and possibly enigmatic) corner of $AdS/CFT$ duality.
%A natural question is how we know that these states
%survive into the large coupling constant regime where the description involves
%gravity. The standard answer is that 

To have a better control in this accounting programme, the microstates are typically constrained to be BPS, which are more
robust under the variations of the string coupling constant and also of other parameters
such as the asymptotic moduli.
However, even this counting is not free from complications. In particular, one
needs to make sure to exclude certain BPS states which can combine into long
representations or decay into BPS constituents. For example, for the case of $su(1,1|1)$ superconformal algebra
\begin{align}\label{algvir}
	{} [ L_m, L_n]&=(m-n) L_{m+n}& & & & \\
	{} [L_0,\calg_{\pm \half}]&=\mp\frac{1}{2}\calg_{\pm \half}\label{rl}, & [L_{\mp 1},\calg_{\pm \half}]&=\pm \calg_{\mp \half}, & 
	[R,\calg_{\pm \half}]&=\calg_{\pm \half}\\
	\{\calg_{\pm \half},\calg_{\pm \half}^\dagger\}&=2L_0\pm R\label{BPS}, &
	\{\calg_{\pm \half},\calg_{\mp \half}^\dagger\}&=2 L_{\pm 1}, &
	\{\calg_{\a },\calg_{\b}\}&= 0,
\end{align} the short spectrum is split into chiral and anti-chiral sectors :
\begin{eqnarray}
	~[h]^{\textnormal{chiral}}_{su(1,1|1)} &=& [(h,2h)]_{sl(2)\oplus u(1)} \oplus [(h+1/2,2h-1)]_{sl(2)\oplus u(1)} \label{chsu11} \\ 
	~[h]^{\textnormal{anti-chiral}}_{su(1,1|1)} &=& [(h,-2h)]_{sl(2)\oplus u(1)} \oplus [(h+1/2,-2h+1)]_{sl(2)\oplus u(1)} , \label{achsu11}
\end{eqnarray} 
where we used the notation $[(sl(2),u(1))]$ for the respective quantum numbers of the bosonic subalgebra $sl(2)\oplus u(1)$, and (\ref{chsu11},\ref{achsu11}) can be combined to obtain special long multiplets $[h]^{L_1},[h]^{L_2}$ : 
\begin{equation}
	[h]^{L_1} := [h]^{\textnormal{chiral}}_{su(1,1|1)} \oplus [h+1/2]^{\textnormal{chiral}}_{su(1,1|1)} \qquad [h]^{L_2} := [h]^{\textnormal{anti-chiral}}_{su(1,1|1)} \oplus [h+1/2)]^{\textnormal{anti-chiral}}_{su(1,1|1)}, 
\end{equation}
whereas a generic long multiplet with $|r|<2h$, is given by (where we drop the $sl(2)\oplus u(1)$ subscript for brevity)
\begin{equation} 
~[(h,r)]^{\textnormal{chiral}}_{su(1,1|1)} = [(h,r)] \oplus [(h+1/2,r-1)] \oplus [(h,r+1)] \oplus [(h+1,r)].
\end{equation} 
This algebra is the unique superalgebra of $\caln=2$ superconformal quantum mechanics, which is the main subject of this note. In particular, the main interest is to be able to do the BPS counting in terms of the $su(1,1|1)$ unitary lowest weight irreducible representations (\ref{chsu11},\ref{achsu11}) for a given such model, and even more optimistically to read the full BPS spectrum\footnote{A partial progress in this direction is achieved for $\caln=(4,4)$ models through a detailed study of $osp(4^*|4)$ representation theory \cite{Dorey:2018klg},\cite{Barns-Graham:2018xdd}.}. $su(1,1|1)$ superconformal indices \cite{Raeymaekers:2024usy} 
\begin{equation}
\cali_\pm (\zeta) = \tr \left( (-1)^F e^{-\beta \calh_\pm }\zeta^{\pm J} \right) , \label{ind}
\end{equation}	 
serve precisely for this purpose, which we compute in Section \ref{sectn2main} in a slightly more general setting than \cite{Raeymaekers:2024usy}, namely by allowing for the presence of non-isolated fixed points, i.e for fixed point submanifolds. 
 
Conformal invariance for one-dimensional sigma models requires that the target space has a conical geometry \cite{Michelson:1999zf},\cite{Gibbons:1998xa}, and hence in superconformal quantum mechanics\footnote{See \cite{Fedoruk:2011aa}, \cite{Britto-Pacumio:2000fnm} for reviews on superconformal mechanics, and \cite{deAlfaro:1976vlx},\cite{Fubini:1984hf} for the original works.} we have two complications being non-compactness and singularity of our target space cones, which are generally not considered in the applications of equivariant localization theorems. The first issue is resolved by considering the spectrum of $L_0$ instead of the original Hamiltonian $H$, which has an effect of introducing a harmonic potential given by the special conformal charge $K\sim r^2$, and hence acting as an IR-regulator. At the level of the superconformal algebra, this is realized through a similarity transformation such that the discrete spectrum of the dilatation operator %(which is discrete due to different inner product used in superconformal quantum mechanics compared to standard supersymmetric quantum mechanics \cite{Dorey:2022ics}) 
maps to that of $L_0$. The second issue however is more serious and \textit{always}\footnote{Except from the `flat' $\mathbb{C}^{k}$ models, which are examined in full detail via various approaches in \cite{Raeymaekers:2024usy}.} exists, since any cone is singular at least at a point $r=0$, the tip of the cone, where the curvature does not vanish, but rather blows up. Hence, even the wave-functions corresponding to eigenstates of $L_0$ will be ill defined at this point, and so is a counting done by an index localizing to this singular point. We overcome this problem in two steps. The first step is to further modify our spectral problem by instead counting the wave-functions corresponding to the eigenstates of $L_0 \pm R$, which introduces an auxiliary magnetic background $\cala^\pm$ to the original problem, and breaks the conformal invariance. Since both $L_0$ and $R$ have a discrete spectrum (in terms of $su(1,1|1)$ irreps), this maps the counting problem to the \textit{unitary} ground states of $L_0 \pm R$, as the BPS bound is a unitarity bound given by $2h\geq |r|$. The second and key point is that (\ref{ind}) is now computed as an equivariant Witten index \cite{Witten:1982df} for a model which is \textit{not} conformally invariant, and hence the fixed point locus is determined by an `arbitrary' Reeb-like vector $\rho^A$ (up to (\ref{rsym})), rather than the conformal one (\ref{confcond2}), and accordingly the refinement in (\ref{ind}) tracks through arbitrary smooth supersymmetry and global isometry preserving deformations of the metric. Concretely, given any SCQM ($H$) defined on a KT (K\"ahler with torsion) space $X$ (with a fixed complex structure), with a singular metric $G$ and the conformal Reeb vector fixed by the homothety, (\ref{ind}) computes the BPS spectrum of SQM ($\calh_\pm$) defined on the same space $X$ with a smooth metric\footnote{The resolved metric $\tilde{G}$ must asymptotically agree with the singular metric $G$ as the resolution parameter goes to zero. In the case of ADHM quiver mechanics, where the target space can be realized as a Hyper-K\"ahler quotient, this resolution parameter coincides with the FI parameter, and has the effect of lifting the singular isolated fixed point locus from the tip of the cone to regular points on the resolved space \cite{Barns-Graham:2018xdd},\cite{Dorey:2022ics},\cite{Aharony:1997an}.} %$^,$\footnote{CŞ would like to thank Andy Zhao for clarifying this point.} 
$\tilde{G}$ and a generically different Reeb vector determined by the holomorphic global isometry $J$, assuming that $(X,G)$ possesses such smooth resolution. This interpretation %for resolving the singularity
is similar to the approach applied successfully for arbitrary quasi-regular K\"ahler cones by Martelli, Sparks, and Yau \cite{Martelli:2006yb}, whereas the relevance for superconformal quantum mechanics was observed in \cite{Dorey:2019kaf}.

One physical motivation originates from a subset of solutions to $\caln=2$, $d=4$ supergravity
obtained from CY compactification of Type-II string theory, where one obtains two
types of BPS solutions: single-centered, or multi-centered. The moduli space of the multi-centered ones is parameterized by Denef equation \cite{Denef:2000nb} :
\begin{equation}
	|\calz_Q| \sin\left( \alpha_Q -\alpha \right)|_{r= \infty} = \sum^n_{p=1} \frac{\left<\Gamma_p,\Gamma_Q\right>}{2 |\vec{x}_p-\vec{x}_Q|} , \label{denef4d}
\end{equation}
which at the same time gives a simple
physical halo-like configuration for a classical BPS bound state. For such bound states with a non-vanishing intersection product one derives from (\ref{denef4d}) that the bound state radius becomes
infinite at the wall of marginal stability, that is to say this class of BPS bound states typically exists at only one side of the wall, unlike the single centered black holes which
can exist for any given value of asymptotic moduli. In particular, a BPS index constructed as a `second helicity supertrace' \cite{Dabholkar:2005by}, to count D6+D2-D0 halo states was studied in \cite{Denef:2007vg} :
\begin{equation}
	\Omega^{J_L}_Q = \sum_{J_R} (-1)^{2J_R} N_Q^{J_L,J_R},
\end{equation} where $N_Q^{J_L,J_R}$ is the dimension of the moduli space of $D2$ brane of charge $Q$ and $(j_L,j_R)$-charge under $SU(2)_L \times SU(2)_R$ R-symmetry.

For a comprehensive understanding of BPS bound states, it is useful to consider another low energy effective description, which allows for an explicit count of generic microstates. This is provided by N=4 supersymmetric quiver quantum mechanics \cite{Denef:2002ru} where the wrapped D-branes appear as particles moving in Minkowski space. The interesting point is that one has both Higgs
and Coulomb branches in this quiver mechanics. Concretely, starting from the
supergravity description and taking $g_S \rightarrow 0$ limit one first arrives the Coluomb
branch quivers consisting of multi-centered particles with a bound state radius that
can be mapped to the one in the supergravity description, and lowering it further one
obtains a more stringy Higgs-branch picture. It has been shown before that \textit{generic} multi-centered BPS molecules of the Coulomb branch make only a sub-leading contribution to the black hole entropy \cite{deBoer:2008fk}, whereas there exists a subset of Higgs-branch solutions which form the exponential majority \cite{Bena:2012hf}. It remains however unclear what the physical interpretation of these pure-Higgs states is when the gravitational coupling constant is not vanishingly small.

This Coulomb branch effective quiver mechanics is originally described in terms of (3,4,1)\footnote{The notation denotes number of (bosonic,fermionic,auxiliary) degrees of freedom.} component form by \cite{Smilga:1986rb}, \cite{Denef:2002ru}
\begin{eqnarray}
	L &=& -f_a D^a - U_a D^a + A_{ia} \dot{x}^{ia} + \partial_{ib} U_a \bar{\lambda}^a \sigma_i \lambda^b + \frac{1}{2} G_{ab} \left( \dot{x}^{ia} \dot{x}^{ib} + D^a D^b + i \left( \bar{\lambda}^a \dot{\lambda}^b - \dot{\lambda}^b \lambda^a \right) \right) \nonumber \\ &-& \frac{1}{2} \partial_{ic} G_{ab} \left( \bar{\lambda}^a \sigma_i \lambda^b D^c + \epsilon_{ijk} \bar{\lambda}^a \sigma_j \lambda^b \dot{x}^{kc} \right) - \frac{1}{8} \partial_{jc} \partial_{jd} G_{ab} \lambda^a \lambda^b \bar{\lambda}^c \bar{\lambda}^d , \label{341}
\end{eqnarray} where the background scalar and gauge potentials are given as
\begin{equation}
	U_a = \sum_{b,b\neq a} = \frac{\kappa_{ab}}{2r_{ab}} \qquad\qquad A_{ia} = - \sum_{b,b\neq a} \kappa_{ab} \frac{\epsilon_{ijk} n^j x_{ab}^k}{2r \left( x_{ab}^l n^l -r\right)}, \label{qqmpot}
\end{equation} where $\kappa_{ab}$, $\mu_{ab}$, $f_a$ are respectively DSZ product, mass, FI parameter, and the metric can be written as\footnote{Explicit form of the function $H(x)$, which is not important for our purpose (since $\mu_{ab}=0$ for scaling solutions), can be obtained from \cite{Mirfendereski:2020rrk}.}
\begin{equation}
	G_{ab} = \delta_{ab} \left(\sum_{c,c\neq a} \frac{|\kappa_{ac}|}{4r_{ac}^3}\right) - \frac{|\kappa_{ab}|}{4r_{ab}^3} + \mu_{ab} H(x).
\end{equation}  

Any $\caln=4B$ supersymmetric mechanics described in terms of $(3,4,1)$ multiplets can be rewritten through the gauging of the auxiliary field \cite{Delduc:2006yp},\cite{Delduc:2006pg} in terms of the so called root multiplet $(4,4,0)$ \cite{Bellucci:2005xn}, which then takes a general form \cite{Mirfendereski:2022omg} : 
\begin{equation}
L = \frac{1}{2} G_{AB} D_t x^A D_t x^B + A_A D_t x^A - \frac{i}{2} F_{AB} \chi^A \chi^B + \frac{i}{2} G_{AB} \chi^A \hat{D}_t \chi^B - \frac{1}{12} \partial_{[A} C_{BCD]} \chi^A \chi^B \chi^C \chi^D, \label{440}
\end{equation} and hence in particular the effective Coulomb branch description (\ref{341}) can be brought into this form \cite{Mirfendereski:2020rrk}. $\caln=4$ supersymmetry with $SU(2)_L \times SU(2)_R$ R-symmetry restricts the target space geometry of (\ref{440}) to be Hyper-K\"ahler with Torsion (HKT).

At a generic point of the effective Coulomb branch, the first equation in (\ref{qqmpot}) is equivalent to (\ref{denef4d}), which shows the correspondence between BPS bound states in $d=4$ supergravity and $d=1$ quiver descriptions \cite{Denef:2002ru}. There is a subset of these bound state solutions known as scaling solutions \cite{Denef:2002ru},\cite{Denef:2007vg} captured in a certain scaling limit of (\ref{341}) \cite{Anninos:2013nra},\cite{Mirfendereski:2020rrk}. This has the net effect of putting mass and FI couplings to zero, while the action remains finite and develops a $D(2,1;0)$ superconformal symmetry. In this limit, the root formulation (\ref{440}) takes the form of a gauged superconformal mechanics \cite{Mirfendereski:2022omg} with the target geometry restricted by a set of conformal constraints which can be interpreted as a deformed version of the well-known standard constraints \cite{Michelson:1999zf},\cite{Papadopoulos:2000ka} for conformal invariance. We revisit this gauged superconformal mechanics in Section \ref{sectquiv} particularly specializing to scaling quivers.% We review this gauged formulation of both scaling and non-scaling quiver bound states in explicit term in Section \ref{sectgagqu}.

The root form of scaling quiver mechanics is especially useful for quantization since it gives a geometric description for the supersymmetric ground states. Moreover, it brings a possibility of an explicit count of these states as $D(2,1;0)$ irreducible representations via corresponding superconformal indices \cite{Gaiotto:2004pc} :
\begin{equation}
\tr \left( (-1)^{2J_L^3} y^{L_0 \pm J_L} z^{L_0 \pm J_R} \right), \label{indd21a}
\end{equation}
which we hope to be able to compute in a future work, in particular for scaling quivers. For a similar, yet simpler $\caln=2B$ superconformal quantum mechanics, this is recently achieved \cite{Raeymaekers:2024usy} for the indices (\ref{ind}) under the assumption of a resolved target, and with a fixed point locus consisting of a single isolated point. While we keep the first assumption, in this note we generalize that computation to the presence of a fixed point submanifold, determined by the holomorphic Killing vector field $\rho^A$ of the resolved space\footnote{Indeed the superconformal index of the analog type-A models for various examples (remarkably for the case of generic toric Calabi-Yau 3-folds \cite{Dorey:2019kaf}) can be computed as a regular Dolbeault cohomology on the resolved space, and the fixed point data of the resolved space is sufficient.}. %In the last part, we review the root (\ref{440}) formulation of Coulomb branch quiver mechanics, and % and provide some more details. %and provide somewhat more details on the supersymmetric localization computation, and point out the differences compared to the 
%we review the explicit computation of an analog $su(1,1|1)$ superconformal index. 
	
%Finally, our gauged superconformal quivers fits into a rather small sub-class of more general $\caln=4$ gauged superconformal mechanics, of which geometric and algebraic constraints are also reviewed here, in addition to the less restricted $\caln=2,1$ cases. In the first part of this note we will mainly introduce the necessary technology to formulate scaling quivers from the point of view of gauged superconformal mechanics, while restricting ourselves to the $\caln=2B$ superconformal symmetry, where we also construct and compute the superconformal index via supersymmetric localization. In the second part, generalizing to gauged $\caln=4B$ superconformal symmetry, we will explicitly study scaling quivers from this perspective and revisit the relevant representation theory as a first step towards an analog superconformal index computation.

\section{$\mathcal{N}=2B$ superconformal quantum mechanics} \label{sectn2main}

On a complex topological space $\calm$, we consider $\caln=(0,2)$ supersymmetric non-linear sigma models with $N$-many chiral multiplets. Each such multiplet contains two bosonic, two fermionic real degrees of freedom and no auxiliary degree of freedom. These non-linear sigma models can generically be defined on a family of different Hermitian metrics $\{G_{AB}\}$; $A=1,\cdots,2N$. In this case, since we are considering $(0,2)$ models but not $(1,1)$ models, we also have a complex structure $J$, and $\caln=2$ supersymmetry restricts the geometry $(G,J)$ with the constraints : 
\begin{equation}
	G_{AC} J^C{}_B + G_{BC} J^C{}_A = 0 \qquad J^A{}_C J^C{}_B = - \delta^A_B \qquad \caln(J)^A{}_{BC} = 0 \qquad \hat{\nabla}_{(A} J_{B)C} = 0. \label{jconst}
\end{equation}    
In addition to $\caln=2$ supersymmetry we also demand $u(1)_R$ R-symmetry transforming the coordinates as 
\begin{equation}
	\delta_r x^A = -2r \rho^A \qquad\qquad \delta_r \chi^A = -2r \left(\partial_B \rho^A +  \frac{1}{2} J^A{}_B \right) \chi^B , \label{u1rsym}
\end{equation} for some vector field $\rho^A$. This restricts the last constraint in (\ref{jconst}) slightly further : 
\begin{equation}
	\hat{\nabla}_A J_{BC} = 0 , \label{rsym1}
\end{equation}
and requires the extra constraints : 
\begin{equation}
	L_\rho G_{AB} = 0 \qquad\qquad L_\rho J^A{}_B = 0 . \label{rsym}
\end{equation}
These models become also conformally invariant if there exists a holomorphic closed homothety $\xi$ such that the refined geometry $(G,J,\xi)$ further satisfies the constraints 
\begin{equation}
	L_\xi G_{AB} = - G_{AB} \qquad\qquad L_\xi J^A{}_B = 0 \qquad\qquad \xi_A = -\frac{1}{2} \partial_A K, \label{confcond1}
\end{equation} which are restrictive enough to fix the vector field $\rho$ as 
\begin{equation}
	\rho^A_{\textnormal{conformal}} = -J^A{}_B \xi^B . \label{confcond2}
\end{equation} Hence, \textit{only} for $\caln=2$ non-linear sigma models which are additionally conformally invariant, (\ref{rsym}) becomes a consequence of the conformal symmetry conditions (\ref{confcond1},\ref{confcond2}). Such metrics $G$ satisfying (\ref{confcond1}) are always\footnote{Except for the flat geometries.} singular at least at a single point $\{0\}$, since the condition 
\begin{equation}
	\nabla_A \xi^B = -\frac{1}{2} \delta^A_B, \label{xi12}
\end{equation} implied by (\ref{confcond1}), itself implies \cite{Gibbons:1998xa} that the conformal metric always takes the form of a cone metric 
\begin{equation}
	G = dr^2 + r^2 g_{ij}(\{x\}) dx^i dx^j, \label{conemetrc}
\end{equation} where $\{0\}$ corresponds to the tip of the cone.

This non-linear sigma model for $N$ chiral multiplets can be obtained easily from the $\caln=1$ superspace action \cite{Coles:1990hr} as 
\begin{equation}
	\call = \frac{1}{2} G_{AB} \dot{x}^A \dot{x}^B + A_A \dot{x}^A - \frac{i}{2} F_{AB} \chi^A \chi^B + \frac{i}{2} G_{AB} \chi^A \left(\dot{\chi}^B + \hat{\Gamma}^B{}_{CD} \dot{x}^C \chi^D \right) - \frac{1}{12} \partial_{[A} C_{BCD]} \chi^A \chi^B \chi^C \chi^D, \label{genn2chrv}
\end{equation} where in general one can pick $\hat{\Gamma}$ as to be slightly more general\footnote{In that case then the extra part $\mathcal{B}_{ABC}$ should necessarily contain a symmetric part because Bismut is the unique totally-antisymmetric one. In case one chooses to work with this slightly more general torsion tensor than the Bismut one, then there is an additional constraint $J^B{}_{[E} \partial_A C_{BCD]} = 0$, which becomes automatic in the Bismut case due to (\ref{bismtors}).} than the Bismut one, but we work with this special choice, i.e. 
\begin{equation}
	\hat{\Gamma}^A{}_{BC} = \Gamma^A{}_{BC} + \frac{1}{2} C^A{}_{BC}, 
\end{equation} where the torsion $C^A{}_{BC}$ is accordingly fixed due to (\ref{rsym1}) by the complex structure $J^A{}_B$ as
\begin{equation}
	C_{ABC} = J_A{}^D J_B{}^E J_C{}^F \left( \nabla_D J_{EF} + \nabla_E J_{FD} + \nabla_F J_{DE} \right), \label{bismtors}
\end{equation}
which should further satisfy 
\begin{equation}
	L_\rho C_{ABC} = 0 , \label{rsymcnd1}
\end{equation} for the invariance of (\ref{genn2chrv}) under R-symmetry. Moreover, the target space geometry becomes K\"ahler if 
\begin{equation}
	\rho^A C_{ABC} = 0. \label{rsymcnd2}
\end{equation} 

The conformal invariance further requires 
\begin{equation}
	L_\xi C_{ABC} = - C_{ABC} \qquad\qquad \xi^A C_{ABC} = 0, \label{confconds2}
\end{equation} which together with (\ref{confcond2}) implies (\ref{rsymcnd1}). Now via (\ref{rsym1},\ref{xi12}) and the second constraint in (\ref{confconds2}), we obtain a nice identity
\begin{equation}
	\rho^C C_{CAB} = J_{AB} + 2 \nabla_A \rho_B. \label{niceid}
\end{equation}

Finally, existence of the background gauge-field $A_A$, with the corresponding field strength $F=dA$, enhances the previous set of R-symmetry constraints to 
\begin{equation}
	F_{AC} J^C{}_B + F_{CB} J^C{}_A = 0 \qquad\qquad i_\rho F  = 0, \label{rsymgf}
\end{equation}  whereas conformal symmetry requires 
\begin{equation}
	i_\xi F = 0,
\end{equation} which together with (\ref{confcond2}) implies the second condition in (\ref{rsymgf}).

The corresponding superconformal symmetry algebra realized by (\ref{genn2chrv}) is $su(1,1|1)$ \cite{Michelson:1999zf}, \cite{Papadopoulos:2000ka}, which we prefer to express in a basis such that the complex supercharges $\calg_{\pm \half}$ satisfy\footnote{More properly, we define a similarity transformation which maps the dilatation generator to $L_0 = \frac{1}{2\omega} \left(H+ \omega^2 K \right)$, so that the supercharges $\calg_{\pm \half}$ satisfy $\{\calg_{\pm \half},\calg_{\pm \half}^\dagger\}= \omega \left( 2L_0 \pm R\right)$. We then choose to work in the units such that $\omega =1$.}  
\begin{equation}
	\{ \calg_{\pm \half}, \calg_{\pm \half}^\dagger \} = 2L_0 \pm R \qquad\qquad [R,\calg_{\pm \half}] = \calg_{\pm \half},
\end{equation} where $L_0 = \frac{1}{2} (H+K)$, and $H,K,R$ are respectively the Hamiltonian, special conformal charge, and the $R$-charge. These fermionic charges $\calg_{\pm \half}$ are constructed by combining supercharges and conformal supercharges of the parent superconformal model (\ref{genn2chrv}), which we now review. From \cite{Mirfendereski:2022omg} we recall that the supercharges $Q^\alpha$ and conformal supercharges $S^\alpha$ are given by
\begin{align}
	Q^1 =& - \chi^A J_{A}^{\ B} \P_B  + {i \over 2}  J_{[A}^{\ \ D} C_{BC]D} \chi^A \chi^B \chi^C, &   	S^1 =&    2  \chi^A J_{AB} \xi^B    \label{GNC1}\\
	Q^2 =&   \chi^A \P_A - {i \over 6} C_{ABC} \chi^A \chi^B \chi^C, &  S^2 =&  - 2 \chi^A  \x_{ A} ,
	\label{GNC2}
\end{align} with
\begin{equation}
	\P_A = \tilde  p_A - A_A  - {i \over 2} \left( \o_{ABC} - \half C_{ABC} \right) \chi^B \chi^C.
\end{equation} The generators $\calg_{\pm \half}$ are then defined as 
\begin{equation}
	\calg_{\pm \half} = \calq \mp i S, \qquad\quad \calq= \frac{1}{2}(Q^1+iQ^2) \qquad \cals = \frac{1}{2} (S^1 + iS^2),
\end{equation} which explicitly give 
\begin{equation}
	\calg_{\pm \half} = \frac{1}{2} \left(\delta^B_A + i J^B_A\right)\chi^A \left(\Pi_B - \cala_B^\pm\right) + \frac{1}{12} \left(\delta^D_A + 3i J_{[A}{}^D \right)C_{BC]D} \chi^A \chi^B \chi^C, \qquad \cala_A^\pm = \mp 2\rho_A.
\end{equation}
%Explicit forms of some of these charges are given in App-\ref{appsusies}. 

Remaining superconformal charges of (\ref{genn2chrv}) are given as \cite{Michelson:1999zf} (in the conventions of \cite{Mirfendereski:2022omg}\footnote{with $R=-2R_{\textnormal{there}}$})
\begin{eqnarray}
	H &=& \frac{1}{2} G_{AB} \Pi^A \Pi^B + \frac{i}{2} F_{AB} \chi^A \chi^B + \frac{1}{12} \partial_{[A} C_{BCD]} \chi^A \chi^B \chi^C \chi^D \\
	D &=& \xi^A \Pi_A \\
	K &=& 2 \xi^A \xi_A \\
	R &=& 2 \rho^A \Pi_A - 2i \nabla_A\rho_B \chi^A \chi^B .
\end{eqnarray}

On the spectrum of the models (\ref{genn2chrv}), there exists a well defined fermion number $F$ which satisfies \cite{Raeymaekers:2024usy} 
\begin{equation}
	~[F,\calg_{\pm \half}] = F,   \qquad F = \frac{i}{2} J_{AB} \chi^A \chi^B + \frac{N}{2},
\end{equation}  and hence 
\begin{equation}
	~[J,\calg_{\pm \half}] = 0 \qquad\qquad J := R-F+c,
\end{equation} extending the superconformal algebra to $su(1,1|1) \oplus u(1)_J$. This also provides the classical expression for the $J$-charge
\begin{eqnarray}
	J &=& -2 \rho^A \Pi_A + 2i \nabla_A\rho_B \chi^A \chi^B \nonumber \\ &=& -2 \rho^A \left( \tilde{p}_A - A_A - \frac{i}{2} \omega_{ABC} \chi^B \chi^C \right) + i \nabla_A\rho_B \chi^A \chi^B - \frac{i}{2} \rho^C C_{CAB} \chi^A \chi^B,
\end{eqnarray} where we used the identity (\ref{niceid}).
The action of this isometry $u(1)_J$ on the coordinates is given by (cfr. (\ref{u1rsym}))
\begin{equation}
	\delta_\epsilon x^A = \epsilon \rho^A \qquad\qquad \delta_\epsilon \chi^A = \epsilon \partial_B \rho^A \chi^B.  \label{jtransf}
\end{equation}
So, a character-valued index \cite{Goodman:1986wq} with respect to this global isometry is introduced \cite{Raeymaekers:2024usy} as 
\begin{equation}
	\mathcal{I}^\pm_\lambda = Tr \left[ (-1)^F e^{-\beta (H+K \pm R) } \zeta^{\pm J} \right] = Tr \left[ (-1)^F e^{-\beta \mathcal{H}^\pm_\lambda}\right], \label{indchar}  
\end{equation} where we defined 
\begin{equation}
	\mathcal{H}^\pm_\lambda  = H + K \pm R  \mp i \lambda J \qquad\qquad \zeta := e^{\beta (i\lambda)}\,,\;i\lambda \in \mathbb{R}. \label{href}
\end{equation}

We compute this index in the standard manner via supersymmetry localization à la Álvarez-Gaumé \cite{Alvarez-Gaume:1983zxc} for the corresponding real supercharge (which corresponds to torsionful Dirac operator \cite{Raeymaekers:2024usy})
\begin{equation}
	D := \calg_{\pm \half} - \calg_{\pm \half}^\dagger, \label{dsusy}
\end{equation}
and obtain that % the 1-loop contribution 
it is given by the formula 
\begin{eqnarray}
	&&\mathcal{I}^\lambda_{\pm} = i^N \int_{\calm_0} \prod_{m=1}^{\dim \calm_0} dx_0^m d\eta_0^m  \frac{\exp \left( \frac{i}{2}\omega^{\lambda}_{AB} \eta_0^A \eta_0^B \right)}{\det'{}_{\textnormal{PBC}} \left( - \delta_{AB} \partial_\tau -i \hat{{R}}^{\pm,\lambda}_{AB}\right)^{1/2}} = \int_{\calm_0} ch(\omega^\lambda) \wedge \hat{A}(\hat{R}^{\pm,\lambda});   \nonumber \\ \hspace{1cm} && 	\hat{R}^{\pm,\lambda}_{AB} = {\hat{R}}_{ABCD} \eta_0^C \eta_0^D \mp 2i \lambda \partial_A \rho_B \qquad\qquad 
	\omega^\lambda_{AB} = \tilde{F}^\pm_{AB} +  i \frac{\lambda}{2} \calf^\pm_{AB},  \label{res3}
\end{eqnarray}  where 
\begin{equation}
	\calf^\pm =d\cala^\pm = \mp 4\nabla_{[A}\rho_{B]} \label{auxpot}
\end{equation}
denotes the auxiliary potential. This result for the index is equivalent to Niemi-Tirkkonen equivariant localization theorem \cite{Niemi:1992nt},\cite{Niemi:1993ia} as expected (since $D$ corresponds to Dirac operator). To evaluate this formula for singular conical geometries of various $\caln=2B$ superconformal quantum mechanical non-linear sigma models, one has to find the fixed point locus $\calm_0$ on the corresponding resolved space, which is given by\footnote{In general we also have time-dependent classical vacua (instanton) configurations given by $\dot{x}^A = \mp 2i \lambda \rho^A$, which we do not consider.} 
\begin{equation}
	D \chi = 0 ~~ \Rightarrow ~~ \rho^A= 0,
\end{equation} where the vector field $\rho^A$ is determined by the conditions (\ref{rsym}). Moreover, since the inverse Legendre transform $\mathcal{L}_\lambda$ of the `Hamiltonian' $\mathcal{H}_\lambda$ is not conformally invariant, $\rho^A$ is in general different than the conformal one (\ref{confcond2}), since on the resolved space there exists no homothety. 

In the remaining parts of this section, we give more details for the supersymmetry localization computation of (\ref{indchar}), starting with the unrefined case $\zeta =1$, for general $\caln=2B$ superconformal mechanics on resolved targets.

\subsection{Superconformal index as an index for auxiliary quantum mechanics}

The path integral for the index
\begin{equation}
	\mathcal{I}_\pm=Tr\left[ (-1)^F e^{-\beta (H+K \pm R)} \right] = \int [D\chi] [Dx] e^{-\beta \int_0^\beta d\tau \call_\pm^E} \label{sconfinddef}
\end{equation} involves the Lagrangian $\call_\pm^E$ which corresponds (after Wick rotation) to the inverse Legendre transform of $\{\calg_\pm,\calg^\dagger_\pm\}$, and is related to the vanilla model (\ref{genn2chrv}) by a simple shift of the background gauge potential, i.e.\footnote{Note that $\xi^2$ term cancels with $K$ in the derivation of (\ref{lpmlag}).}
\begin{equation}
	\call_\pm = \call [A \rightarrow \tilde{A}_\pm = A + \cala_\pm] , \label{lpmlag}
\end{equation}  which is invariant under $\caln=2$ supersymmetries :
\begin{eqnarray}
	&&	\delta_{\calg^\dagger_{\pm 1/2}} x^A = - \frac{i\epsilon}{2} (J^A{}_B -i \delta^A_B) \chi^B  \label{gdag1u} \\ &&
	\delta_{\calg^\dagger_{\pm 1/2}} \chi^A = -\frac{\epsilon}{2} \left(J^A{}_B +i\delta^A_B  \right)\dot{x}^B + \frac{i\epsilon}{2} \partial_C J^A{}_B \chi^C \chi^B \label{gdag2u} \\ &&
	\delta_{(\calg_{\pm 1/2} - \calg^\dagger_{\pm 1/2})}  x^A = \epsilon \chi^A \qquad\qquad 
	\delta_{(\calg_{\pm 1/2} - \calg^\dagger_{\pm 1/2})} \chi^A = i\epsilon \dot{x}^A_,  \label{ddpmru}
\end{eqnarray} as well as under the $u(1)_J$ transformation (\ref{jtransf}).

%Note that as far as the computation of (\ref{sconfinddef}) is concerned, we can deform the auxiliary model $\call_\pm$ by an additional supersymmetry exact term \cite{Alvarez-Gaume:1983uye} :
%\begin{eqnarray}
%	\call_\pm \rightarrow \tilde{\call}_\pm &\equiv& \call_\pm \mp 2i  \delta(\rho_A \chi^A) \\ &=& \call_\pm \pm 2 \left(\rho_A \dot{x}^A - i\nabla_A \rho_B \chi^A \chi^B  \right),
%\end{eqnarray} where $\delta$ denotes (\ref{ddpmru}) with the infinitesimal parameter set to $\epsilon = 1$. We then get via (\ref{lpmlag})
%\begin{equation}
%	\tilde{\call}_\pm = \call_\pm - \left( \cala^\pm_A \dot{x}^A - \frac{i}{2} \calf^\pm_{AB} \chi^A \chi^B \right) = \call ,
%\end{equation} i.e. this $\delta$-exact deformation removes the shift in the background potential and recovers back the original vanilla model (\ref{genn2chrv}). Indeed, 

We note that when $\call_\pm \rightarrow \call$, and the torsion is put to zero, the index is manifestly equivalent to the index of Dirac-operator \cite{Ivanov:2010ki},\cite{Smilga:2011ik}, first computed by Atiyah and Singer \cite{Atiyah:1968mp},\cite{Atiyah:1971rm}, and then by Álvarez-Gaumé \cite{Alvarez-Gaume:1983zxc},\cite{Alvarez-Gaume:1986ggp} and Friedan and Windey \cite{Friedan:1983xr} via supersymmetry path integral, except that in our case we have a noncompact target. For simplicity we will assume that the K\"ahler form $\Omega$ satisfies the condition
\begin{equation}
	\partial \bar{\partial} \Omega =0,
\end{equation} so that the four-fermion term in (\ref{genn2chrv}) drops out and thus we are able to apply standard supersymmetry localization \cite{Smilga:2011ik}.  Such target geometries are called strong K\"ahler with torsion (SKT), but our results straightforwardly generalize to weak K\"ahler with torsion (wKT) geometries where the derivative of the torsion tensor is non-vanishing. However, this brings a simplification rather than a complication because in the wKT cases, the whole contribution from the torsion can be made vanishing through a continuous deformation while preserving the $\caln=2$ supersymmetry of (\ref{genn2chrv}) in computing the index \cite{Smilga:2011ik}. 

Let us now summarize the computation of (\ref{sconfinddef}) through explicit supersymmetry localization for SKT geometries. First, we note that the supersymmetry generator (\ref{dsusy}) acts on $\call_\pm$ as given in (\ref{ddpmru}),
%\begin{equation}
%	\delta_{(\calg_{\pm \half} - \calg_{\pm \half}^\dagger)} x^A = \epsilon \chi^A \qquad\qquad \delta_{(\calg_{\pm \half} - \calg_{\pm \half}^\dagger)} \chi^A = i \epsilon \dot{x}^A , \label{susystrnsf}
%\end{equation}   
and hence we observe that a supersymmetry-exact generalization of (\ref{genn2chrv}) is given by
\begin{equation}
	\call_\pm(\kappa) = \kappa \delta_{(\calg_{\pm 1/2} - \calg^\dagger_{\pm 1/2})}\left( - \frac{i}{2} G_{AB} \dot{x}^A \chi^B - \frac{1}{12} C_{ABC} \chi^A \chi^B \chi^C \right) + \tilde{A}^\pm_A \dot{x}^A - \frac{i}{2} \tilde{F}^\pm_{AB} \chi^A \chi^B, \label{unrefdx1}
\end{equation} 
with an arbitrary parameter $\kappa$. Note that (\ref{unrefdx1}) reduces to (\ref{lpmlag}) for $\kappa=1$. Now, expanding (\ref{unrefdx1}) over the fluctuations 
\begin{equation}
	x = x_0 + \frac{\xi}{\sqrt{\kappa}} \qquad\qquad \chi = \eta_0 + \frac{\eta}{\sqrt{\kappa}}, \label{flucts}
\end{equation} and considering the limit $\kappa \rightarrow \infty$, the Euclidean path integral (\ref{sconfinddef}) evaluates to %(\ref{sconfinddef}) of $\check{L}(\kappa)$ localizes onto the fixed point locus 
%\begin{equation}
%	\calm_0 = \{x_0|\rho(x_0)=0\}, \label{fixdpts}
%\end{equation} and gives 
\begin{equation}
	\mathcal{I} =i^{N} \int_{\calm} \prod_{K=1}^{\textnormal{dim} \calm} dx_0^K d\eta_0^K \frac{e^{\frac{i}{2}\tilde{F}^\pm_{AB} \eta_0^A \eta_0^B}}{\det'_{\textnormal{PBC}} \left( - \delta_{AB} \partial_\tau - i \hat{R}_{ABCD} \eta_0^C \eta_0^D \right)^{1/2}} = \int_{\calm} ch(\tilde{F}^\pm) \wedge \hat{A}(\hat{R}), \label{unrfres1}
\end{equation} where $\hat{R}$ is the torsionful Riemann tensor. We obtain the corresponding index for the wKT geometry simply by replacing $\hat{R}$ with the standard torsionless Riemann tensor \cite{Smilga:2011ik}. We also note that it is the `net' field $F+\calf^\pm$ rather than only the background field $F$ that appears inside the Chern character. Up to these differences, (\ref{unrfres1}) coincides with the standard result of Atiyah-Singer index. However, in this case, since $\calm$ is noncompact the result (\ref{unrfres1}) is divergent.% in that the integral is taken only over the subspace $\calm_0$ due to non-standard supersymmetry transformation of fermions (\ref{susystrnsf}).

%As can be noticed (or as will become clear), the unrefined index is actually a waste of resources at our disposal, i.e. although the model $\calh_\pm$ is much less restricted than $H$ as it breaks conformal and $R$ symmetries, Witten index constructed from $\calh_\pm$ does not make a good use of this freedom. 

Finally let us also note that to compute the index (\ref{sconfinddef}), we merely used the $\caln=2$ supersymmetry invariance of the auxiliary Lagrangian $\call^E_\pm$ appearing in the index path integral. In fact, unlike the original superconformal model (\ref{genn2chrv}), $\call_\pm$ is not invariant under $u(1)_R$ and conformal transformations due to the appearance of the extra potential $\cala_\pm$. In other words, Witten indices constructed from $\calh_\pm$ do not make a good use of this freedom.

%This observation that the auxiliary model $\check{L}$ has less symmetry than the original superconformal model, when combined with the standard invariance property of the Witten index under continuous symmetry preserving deformations, will become quite useful for computing a refined index.

%\item Example: Free model, special case of 

\subsection{Supersymmetric Localization of the Refined Index}

We now return to the refined superconformal index (\ref{indchar}), which can be interpreted as the equivariant Witten index for the refined Hamiltonian (\ref{href}), with the $u(1)_J$ isometry generated by a holomorphic Killing vector field $\rho^A$. Crucially, the refined Lagrangian $L^\pm_\lambda$ appearing in the index path integral is not conformally invariant. Hence, only $\caln=2$ supersymmetry and $u(1)_J$ symmetry are preserved in the index path integral, whereas the stringent condition (\ref{confcond2}) on $\rho^A$ does not hold. %As we will argue this freedom turns out to be quite useful. 

The Lagrangian corresponding to the refined Hamiltonian 
\begin{equation}
	H^\pm_{\lambda}  = H + K \pm R \mp i \lambda J,~~i\lambda \in \mathbb{R},   \label{refhamdfn}
\end{equation} can be easily obtained via inverse Legendre transform as
\begin{eqnarray}
	L^\pm_{\lambda}  &=& \frac{1}{2} G_{AB} \dot{x}^A \dot{x}^B + \tilde{A}^\pm_A \dot{x}^A + \frac{i}{2} G_{AB} \chi^A \hat{\nabla}_t \chi^B - \frac{i}{2} \tilde{F}^\pm_{AB} \chi^A \chi^B \mp  \lambda \left(\nabla_A \rho_B + \frac{1}{2} \rho^C C_{CAB} \right)\chi^A \chi^B \nonumber \\ && \hspace{0.cm} \mp 2i\lambda \rho_A \dot{x}^A - 2\lambda^2 \rho_A \rho^A - \frac{1}{12} \partial_{[A} C_{BCD]} \chi^A \chi^B \chi^C \chi^D , \label{llambpmr}
\end{eqnarray} and is invariant under \textit{refined} $\caln=2$ supersymmetries : %which in the Euclidean signature becomes 
%\begin{eqnarray}
%	{L}_{\lambda}^{E} &=& \frac{1}{2} G_{AB} \dot{x}^A \dot{x}^B -i\tilde{A}_A \dot{x}^A + \frac{1}{2} G_{AB} \chi^A \hat{\nabla}_t \chi^B + \frac{i}{2} \tilde{F}_{AB} \chi^A \chi^B +  \lambda \left(\nabla_A \rho_B + \frac{1}{2} \rho^C C_{CAB} \right)\chi^A \chi^B \nonumber \\ && \hspace{0.cm} - 2\lambda \rho_A \dot{x}^A + 2\lambda^2 \rho_A \rho^A + \frac{1}{12} \partial_{[A} C_{BCD]} \chi^A \chi^B \chi^C \chi^D. \label{llambref}
%\end{eqnarray} 
\begin{eqnarray}
	&&	\delta_{\calg^\lambda_{\pm \half}} x^A = -\frac{i\epsilon}{2}  \left( J^A{}_B+i \delta^A_B\right) \chi^B  \label{n2gpm1ref} \\
	&& \delta_{\calg^\lambda_{\pm \half}} \chi^A = -\frac{\epsilon}{2} \left(J^A{}_B -i\delta^A_B  \right) \left( \dot{x}^B \mp  2i\lambda \rho^B \right) + \frac{i\epsilon}{2} \partial_C J^A{}_B \chi^C \chi^B, \label{n2gpm2ref} \\ 
	&& \delta_{\calg^{\lambda\dagger}_{\pm 1/2}} x^A = - \frac{i\epsilon}{2} (J^A{}_B -i \delta^A_B) \chi^B  \label{gdag1ref} \\ 
	&& \delta_{\calg^{\lambda\dagger}_{\pm 1/2}} \chi^A = -\frac{\epsilon}{2} \left(J^A{}_B +i\delta^A_B  \right) \left( \dot{x}^B \mp  2i\lambda\rho^B \right) + \frac{i\epsilon}{2} \partial_C J^A{}_B \chi^C \chi^B. \label{gdag2ref} \\
	&& \delta_{(\calg^\lambda_{\pm 1/2} - \calg^{\lambda\dagger}_{\pm 1/2})}  x^A = \epsilon \chi^A\,, \qquad\qquad 
	\delta_{(\calg^\lambda_{\pm 1/2} - \calg^{\lambda\dagger}_{\pm 1/2})} \chi^A = i\epsilon \left( \dot{x}^A \mp 2i \lambda \rho^A \right).  \label{ddpmrref}
\end{eqnarray}
Comparing these with the unrefined analogs (\ref{gdag1u}-\ref{ddpmru}), we see that the supersymmetry transformations of fermions receive a fugacity-dependent extra contribution determined by the vector $\rho^A$. This leads to localization of the index path integral 
\begin{equation}
	\mathcal{I}_\lambda^\pm = \int [Dx] [D\chi] e^{-\int_0^\beta d\tau L^{\pm,E}_\lambda}, \label{indpiref}
\end{equation} 
to a subspace determined by the fixed point locus $\calm_0$ of $\rho^A$ : 
\begin{equation}
	\calm_0 = \{x_0|\rho(x_0)=0\}, \label{fixdpts}
\end{equation}
instead of an integration over the full target space $\calm$, and hence eventually brings a finite result unlike the unrefined index.  

As similar to before, we find a useful supersymmetry-exact generalization of (\ref{llambpmr}) :
\begin{equation}
	{L}^\pm_{\lambda,\kappa} = \kappa \delta_{\lambda} \left(-\frac{i}{2} G_{AB} \dot{x}^A \chi^B - \frac{1}{12} C_{ABC} \chi^A \chi^B \chi^C \right) %+ \tilde{A}^\pm_A \dot{x}^A 
	- \frac{i}{2} \tilde{F}^\pm_{AB} \chi^A \chi^B \mp  \lambda \nabla_A \rho_B \chi^A \chi^B - 2 \lambda^2 \rho_A\rho^A , \label{dexctref}
\end{equation}
where $\delta_\lambda$ is used as a shorthand for (\ref{ddpmrref}), and we did not write $\mathcal{O}(\dot{x}^A)$ terms which do not contribute to path integral index as $\kappa \rightarrow \infty$.%  and we observe that (\ref{dexctref}) reduces to (\ref{llambpmr}) for $\kappa=1$.%\footnote{More correctly one obtains $L^\pm_{\lambda,\kappa=1}= L^\pm \pm i \lambda \rho_A \dot{x}^A}.

%Note that the transformations (\ref{boostsusy}) are the same as (\ref{susystrnsf}) but with a rescaling factor of $\lambda$ in front of $\rho^A$ describing the conformal supersymmetry of the fermions. Since for the $\caln=2$ supersymmetry invariance (even for the $R$-symmetry) of (\ref{llambpmr}) itself the scaling of $\rho^A$ is arbitrary this is a legitimate move.

We are now ready to use the localization principle for the index path integral (\ref{indpiref}) computed with the Euclidean continuation of (\ref{dexctref}), which after expanding over (\ref{flucts}) up to quadratic order in fluctuations, and considering the limit $\kappa\rightarrow \infty$, gives %the 1-loop contribution as 
(\ref{res3})\footnote{which is valid for a SKT geometry. Similarly as before, one obtains the corresponding result for the wKT geometry simply by replacing the Riemann tensor with the torsionless one.}. %To obtain the full result, we note that the classical action also brings a non-vanishing contribution due to refinement : :%(putting $\beta=1$)
%\begin{equation}
%\int_{\calm_0} \prod_{m=1}^{\textnormal{dim} \calm_0} dx_0^m d\eta_0^m \exp \left( \pm i \lambda \nabla_A \rho_B \eta_0^A \eta_0^B \right),
%\end{equation} which precisely cancels the $\mathcal{O}(\lambda)$ contribution in the numerator of the 1-loop determinant (\ref{res3}). Thus, the full-result $\cali_\lambda$ is simply given by replacing $\omega^\lambda$ by $\tilde{F}$ in (\ref{res3}) :
%\begin{equation}
%	\mathcal{I}^\pm_\lambda = \mathcal{I}_{\pm,\lambda}^{\textnormal{1-loop}} (\omega^\lambda \rightarrow \tilde{F}). \label{finresrft}
%\end{equation} 
For the cases where $\calm_0$ consists of only isolated fixed points $\{x_0\}$, it reduces to a simple form 
\begin{equation}
	\mathcal{I}^\pm_\lambda = i^N \sum_{\{x_0\}} \det_{\textnormal{PBC}}{}' \left(-\delta_{AB} \partial_\tau \mp 2\lambda \partial_A \rho_B{(x_0)} \right)^{-1/2}, \label{refindispts}
\end{equation}
as obtained in \cite{Raeymaekers:2024usy}. Hence, unlike the unrefined index which generically gives an infinite result for noncompact spaces, the refined index gives a finite result (at a fixed charge) since the domain of integration is now restricted to be $\calm_0$, but not over the full target space $\calm$. 

Finally, we note that%\footnote{As before we take the infinitesimal supersymmetry parameter as $\epsilon = \sqrt{\omega}$.} 
\begin{equation}
	\mp \lambda \delta_\lambda \left( \rho_A \chi^A\right) = \mp i\lambda \rho_A\dot{x}^A \mp \lambda \nabla_A\rho_B \chi^A \chi^B - 2 \lambda^2 \rho_A\rho^A,
\end{equation} and thus we can deform $\call_\pm^\lambda$ by a $\delta_\lambda$-exact term as%\footnote{The last equality in (\ref{dfrmlref}) requires the condition (\ref{rsymcnd2}), which we assume to be the case in the following.} 
\begin{equation}
	\call_\pm^\lambda \rightarrow \call_\pm^\lambda \mp \lambda \tau \delta_\lambda(\rho_A \chi^A ), \label{dfrmlref} %= \call^{\lambda \tau}_\pm, \label{dfrmlref}
\end{equation} so that (\ref{dexctref}) generalizes
\begin{equation}
	\call_\pm^{\lambda_\tau,\kappa} = \kappa \delta_{\lambda_\tau} \left(G_{AB} \dot{x}^A \chi^B - \frac{i}{12} C_{ABC} \chi^A \chi^B \chi^C \right) + \tilde{A}_A^\pm \dot{x}^A - \frac{i}{2} \tilde{F}_{AB}^\pm \chi^A \chi^B \mp \lambda_\tau \nabla_A \rho_B \chi^A \chi^B - 2 \lambda_\tau^2\rho_A\rho^A ,
\end{equation}  where we introduced the shorthand 
\begin{equation}
	\lambda_\tau := \tau \lambda,\qquad\qquad \tau \in \mathbb{R}.
\end{equation}
So, we see that the deformation (\ref{dfrmlref}) has the net effect of introducing an arbitrary relabeling factor $\tau$ in front of the fugacity $\lambda$ in the main result (\ref{res3}). %Thus, taking into account this deformation, we can rewrite our final result for the refined index as 
%\begin{eqnarray}
%	&&\mathcal{I}^\pm_{\lambda_\tau} = i^N \int_{\calm_0} \prod_{m=1}^{\dim \calm_0} dx_0^m d\eta_0^m  \frac{\exp \left( \frac{i}{2}\tilde{F}^\pm_{AB} \eta_0^A \eta_0^B \right)}{\det'{}_{\textnormal{PBC}} \left( - \delta_{AB} \partial_\tau -i \hat{{R}}^{\lambda_\tau}_{AB}\right)^{1/2}} = \int_{\calm_0} ch(\tilde{F}^\pm) \wedge \hat{A}(\hat{R}^{\lambda_\tau});   \nonumber \\ \hspace{1cm} && 	\hat{R}^{\lambda_\tau}_{AB} = {\hat{R}}_{ABCD} \eta_0^C \eta_0^D - 2i \lambda_\tau \partial_A \rho_B \qquad\qquad 
%	\tilde{F}^\pm = F + \calf^\pm.  \label{res4}
%\end{eqnarray}
Similarly to the unrefined case, (\ref{res3}) is valid for SKT targets, and the generalization to the wKT cases is simply obtained by replacing the torsionful equivariant Riemann curvature with the torsionless one.

To conclude, the superconformal index defined as an equivariant Witten index on the resolved target space can be obtained from the general localization formula (\ref{res3}), once the fixed point locus of the holomorphic Killing vector $\rho$ on this resolved space is determined.

%\begin{itemize}
%	\item Most general superspace action, papadopoulos-coles
%	\item Chiral part only	
%	\item Geometry and constraints
%	\item Global isometry 
%	\item Representation theory for $su(1,1|1) \oplus u(1)$ 
%	\item Superconformal index as an index for auxiliary quantum mechanics
%	\item Example: Free model
%\end{itemize}

		\section{Gauged Quivers} \label{sectquiv}

%In this section we study the gauged $\caln=4$ supersymmetric quantum mechanics of Coulomb branch quivers starting from the 2-node system and generalizing to $N$-nodes. In each case we then consider the additional scaling limit which brings a full conformal symmetry. 

We now move on to the $\caln=4$ case in order to make a connection with the quiver model of our interest mentioned in the Introduction. Due to $\caln=4B$ supersymmetry we now have two additional complex structures, which we express in the covariant form 
\begin{equation}
	J^\rho = (J^i,\mathds{1}) \qquad\qquad \bar{J}^\rho = (-J^i,\mathds{1}) ,\qquad\qquad i=1,2,3.
\end{equation} This leads to HyperK\"ahler with torsion target spaces for the corresponding sigma models, that can be parametrized by the coordinates
\begin{equation}
	x^A:= x^{\mu a} \qquad\qquad A \equiv \mu a = 1,\cdots,4n,\qquad \mu=(i,4),\,\,a=1,\cdots,n.
\end{equation}
We now assume that there exists a set of global isometries described by $n$-many commuting\footnote{Generalization to non-abelian case is given in \cite{Mirfendereski:2022omg}.} Killing vectors $\{k_a\}$ acting on the bosonic and fermionic coordinates as\footnote{Comparing (\ref{gaugetrnsf}) with (\ref{jtransf}) reveals that one can consider $\caln=2$ gauged models where the refinement generator $J$ is gauged \cite{CSgauged}, and thereby building an explicit connection with \cite{Martelli:2006yb} in superconformal mechanics.} 
\begin{equation}
\delta_ \lambda x^A = \lambda^b k^A_b \qquad\qquad \delta_\lambda \chi^A = \lambda^b \partial_C k_b^A \chi^C. \label{gaugetrnsf}
\end{equation} 
Gauging this global isometry by promoting $\lambda^a \rightarrow \lambda^a(t)$, and correspondingly introducing a worldline valued gauge fields $a^a(t)$, (\ref{genn2chrv}) becomes \cite{Mirfendereski:2022omg}
\begin{equation}
	L = \frac{1}{2} G_{AB} D_t x^A D_t x^B + A_A \dot{x}^A + a^a v_a - \frac{i}{2} F_{AB} \chi^A \chi^B  + \frac{i}{2} G_{AB} \chi^A \hat{D}_t \chi^B - \frac{1}{12} \partial_{[A} C_{BCD]} \chi^A \chi^B \chi^C \chi^D  , \label{4402}
\end{equation} where the gauge covariant derivatives are 
\begin{equation}
D_t x^A = \dot{x}^A - a^b k^A_b \qquad\qquad \hat{D}_t \chi^A  = \dot{\chi}^A + \hat{\Gamma}^A{}_{BC} \dot{x}^B \chi^C + a^b \left( \nabla^A k_{bC} + \frac{1}{2} C^A{}_{CD} k^D_b \right) \chi^C,
\end{equation}
and $v_a(x)$ are arbitrary target space valued potentials satisfying the constraint 
\begin{equation}
i_{k_a} F = dv_a.
\end{equation}

We now specialize to a specific choice of a target space data : 
\begin{eqnarray}
	&&G_{\mu a \nu b} = \delta_{\mu \nu} G_{ab} ,\qquad\quad\,\, G_{ab} = \delta_{ab} \left(\sum_{c,c\neq a} \frac{|\kappa_{ac}|}{4r_{ac}^3}\right) - \frac{|\kappa_{ab}|}{4r_{ab}^3} + \mu_{ab} H(x) %\qquad \partial_{ia} G_{bc} = \partial_{(ia} G_{bc)} 
	\label{quivdt1} \\ && C_{\mu a \nu b \rho c} = \epsilon_{\lambda \mu \nu \rho} \partial_{\lambda a} G_{bc} \label{quivdt2} \\ && (J^i)^{\mu a}{}_{\nu b} = (j^i_+)_{\mu \nu} \delta^a_b, \qquad (j^i_\pm)_{\mu \nu} = \mp \left( \delta_{\mu i} \delta_{\nu 4} - \delta_{\mu 4} \delta_{\nu i} \right) - \epsilon_{i\mu \nu 4} \label{quivdt3} \\ && A_A =A_{\mu a} =(A_{ia},-v_a) =(A_{ia},-f_a-U_a), \label{quivdt4}
\end{eqnarray} which is evidently invariant under 
the isometry
\begin{equation}
	k_a^A = \delta^A_4 \partial_a . \label{gaugsymkil}
\end{equation}
Second, we note that by fixing the gauge such that $x^{4a}$ is constant, (\ref{4402}) becomes the $(3,4,1)$ effective Coulomb quiver mechanics (\ref{341}) : 
\begin{equation}
	L_{(4,4,0)} (D_t x^{4a} = -a^a:=D^a) = L_{(3,4,1)}.
\end{equation}  

When $\mu_{ab}=0,f_a =0$, it was shown \cite{Anninos:2013nra},\cite{Mirfendereski:2020rrk},\cite{Mirfendereski:2022omg} that (\ref{4402}) is invariant under $D(2,1;0)$ action. A detailed analysis of the algebra closure and the derivation of geometric constraints for a general target space data, as well as the corresponding restrictions for the specific choice (\ref{quivdt1}-\ref{quivdt4}), can be found in \cite{Mirfendereski:2022omg}. In the $(4,4,0)$ language, this has the interpretation that the conformal symmetry is realized after a reduction from the HKT space to a subspace which is determined by the $U(1)^n$ gauge symmetry (\ref{gaugetrnsf}) :
\begin{equation}
	M_a = - v_a \approx 0 \Leftrightarrow U_a \approx 0.
\end{equation} 
Gauged formalism allows us instead to work on the covering HKT space, and try to compute the corresponding $D(2,1;0)$ index (\ref{indd21a}) with respect to this gauged model. From the computation in the previous section, we can now see the simple reason why this would probably be a powerful tool for computing the index : because in the gauged description of scaling quiver mechanics (\ref{4402}) the analog $\rho$-vector field takes the form \cite{Mirfendereski:2022omg} :
\begin{equation}
	\rho_i = - \epsilon^{ijk} x^{ja} \partial_{ka} - x^{ia} \partial_{4a},
\end{equation} where the shift due to gauged isometry (\ref{gaugetrnsf}) given by the second term lifts the fixed point locus from the singular tip of the cone of the unresolved conformal target space as $x^{4a} \neq 0$. Indeed, in the gauged model $\rho$ is given by 
\begin{equation}
	\rho = -J \cdot \xi_\perp,
\end{equation} where $\xi_\perp$ is not a homothety but rather related to that with a shift given by the Killing vector (\ref{gaugsymkil}) \cite{Mirfendereski:2022omg}. This suggests \cite{CSgauged} that the gauged superconformal mechanical sigma models might be useful to obtain a more precise description of the fixed point locus $\calm_0$ on resolved targets, which is a key ingredient of the localization formula (\ref{res3}).  

\acknowledgments

We thank Nick Dorey, Paolo Rossi, Andrei Smilga, and Andy Zhao for useful discussions. The work of C\c{S} was supported by the European Union's Horizon Europe programme under grant agreement No. 101109743, project Quivers. The research of J.R. was supported by European Structural and Investment Funds and the Czech Ministry of Education, Youth and Sports (Project FORTE CZ.02.01.01/00/22\_008/0004632).
DVdB was supported by the Bilim Akademisi through a BAGEP award.

\bibliographystyle{ytphys}
\bibliography{refs}	
	
\end{document}